\documentclass[preprint,article,accept,moreauthors,pdftex]{Definitions/mdpi} 

\firstpage{1} 
\pubvolume{1}
\issuenum{1}
\articlenumber{0}
\pubyear{2021}
\copyrightyear{2021}
\history{Received: date; Accepted: date; Published: date}

\usepackage[utf8]{inputenc}
\usepackage{siunitx}
\usepackage{bm}

\Title{Geometrically constrained Skyrmions}

\Author{Swapneel Amit Pathak $^{1}$\orcidB{}  and Riccardo Hertel $^{1}$*\orcidA{}}

\AuthorNames{Swapneel Amit Pathak and Riccardo Hertel}

\address{%
$^{1}$ \quad Universit{\'e} de Strasbourg and CNRS, Institut de Physique et Chimie des Mat{\'e}riaux de Strasbourg, F-67000 Strasbourg, France}

\corres{Correspondence: riccardo.hertel@ipcms.unistra.fr}

\abstract{
Skyrmions are chiral swirling magnetization structures with nanoscale size. These structures have attracted considerable attention due to their topological stability and promising applicability in nanodevices, since they can be displaced with spin-polarized currents. However, for the comprehensive implementation of skyrmions in devices, it is imperative to also attain control over their geometrical position. Here we show that, through thickness modulations introduced in the host material, it is possible to constrain three-dimensional skyrmions to desired regions. We investigate skyrmion structures in rectangular FeGe platelets with micromagnetic finite element element simulations. First, we establish a phase diagram of the minimum-energy magnetic state as a function of the external magnetic field strength and the film thickness. Using this understanding, we generate preferential sites for skyrmions in the material by introducing dot-like ``pockets'' of reduced film thickness. We show that these pockets can serve as pinning centers for the skyrmions, thus making it possible to obtain a geometric control of the skyrmion position. This control allows stabilizing skyrmions at positions and in configurations that they would otherwise not attain. Our findings may have implications for technological applications in which skyrmions are used as units of information that are displaced along racetrack-type shift register devices.
}


\begin{document}

\section{Introduction}
Magnetic skyrmions, predicted by theory almost 30 years ago, have advanced to a central topic of research \cite{finocchio2016magnetic,everschor2018perspective,back20202020} in nanoscale magnetism over the last decade following their experimental observation \cite{muhlbauer_skyrmion_2009,yu2010real}. Their particular topological properties \cite{oike2016interplay}, which impart them high stability and particle-like behavior \cite{iwasaki2013universal,iwasaki2013current,xuan2018nonuniform}, combined with their room-temperature availability \cite{yu_near_2011,boulle2016room}, reduced dimensions \cite{heinze2011spontaneous} and their unique dynamic properties \cite{yu2012skyrmion}, render these magnetic structures promising candidates for future spintronic applications \cite{fert_skyrmions_2013}. Skyrmions are formed in non-centrosymmetric magnetic materials exhibiting a sufficiently strong Dzyaloshinsky-Moriya Interaction (DMI) \cite{muhlbauer_skyrmion_2009,yu2010real,seki2012observation}, {\em i.e.}, an antisymmetric energy term that favors the arrangement of the magnetization in helical spin structures with a specific handedness and spiral period. In extended thin films, the formation and stability of skyrmions depends sensitively on various parameters, such as the strength of an externally applied magnetic field, the film thickness, temperature, and the magnetic history of the sample. Phase diagrams have been reported in the literature \cite{yu_near_2011,rybakov_three-dimensional_2013,rybakov_new_2016,beg_ground_2015}, displaying the parameter ranges within which skyrmions are stable and where they may take different forms. Skyrmions may typically either develop individually or in the form of a hexagonal skyrmion lattice. While the occurrence of individual skyrmions makes them attractive candidates for units of information that can be displaced in a controlled way by spin-polarized electric currents, their spontaneous arrangement in the form of a periodic lattice could be interesting for magnonic applications, where  these point-like magnetic structures could play the role of scattering centers of planar spin waves.
One drawback of possible applications of skyrmions is the difficulty of controlling their position. For instance, if skyrmions are to be used as units of information in race-track type shift-register devices, it is not only necessary to be able to displace them in a controlled way, but also to make sure that they are shifted between well-defined positions along the track. In domain-wall based concepts for race-track memory devices \cite{parkin_magnetic_2008}, which preceded the skyrmion-based variants, this control of the position was typically achieved by inserting indentations (``notches'') into the strips \cite{bedau_angular_2007,bogart_effect_2008}. It was shown that such notches represent preferential sites for domain walls, making it possible to trap domain walls at these specific positions, from which they could only be detached after overcoming a certain depinning energy \cite{garcia-sanchez_depinning_2011}. 
Although the possibility to capture skyrmions at specific sites has been addressed in the case of ultrathin films, a geometric control analogous to the pinning of domain walls at notches does not yet seem to be firmly established for skyrmions. In two-dimensional systems, strategies for the pinning of skyrmions include the insertion of point-like defects \cite{liu_mechanism_2013,hanneken_pinning_2016} or atomic-scale vacancies \cite{muller_capturing_2015}. Remarkably, randomly distributed point defects in ultrathin films have also been reported to have little effect on the current-driven skyrmion dynamics \cite{iwasaki_universal_2013}. Motivated to further the discussion on this topic by addressing the three-dimensional case of ``bulk'' DMI, we use finite-element micromagnetic simulations to study the extent to which a geometric control of the skyrmion position in a thin-film element can be achieved by introducing a patterning in the form of local variations of the film thickness. 
Our simulations show that, by locally lowering the film thickness in sub-micron sized dot-shaped regions, skyrmions can in fact be ``captured'' at these geometrically defined sites. We find that, by using geometrical constrains of this type, skyrmions can be stabilized at positions that they would otherwise not adopt. For instance, these geometric manipulations make it possible to generate regular, square lattices of skyrmions which, apart from exceptional situations \cite{karube_robust_2016}, are not observed naturally in non-centrosymmetric ferromagnets. It is argued that this control of skyrmion positions in magnetic thin films can open up new possibilities for skyrmionic devices as well as for concepts of magnonic metamaterials.

\section{Results}

Before addressing the question of how preferential sites for skyrmions can be generated through nanopatterning, we first investigate, as a preliminary study, the field- and thickness dependence of skyrmionic structures forming in  rectangular FeGe platelets. 

\subsection{Chiral magnetization states in a helimagnetic rectangular platelet}
\begin{figure}[H]
\centering
\includegraphics[width=\textwidth]{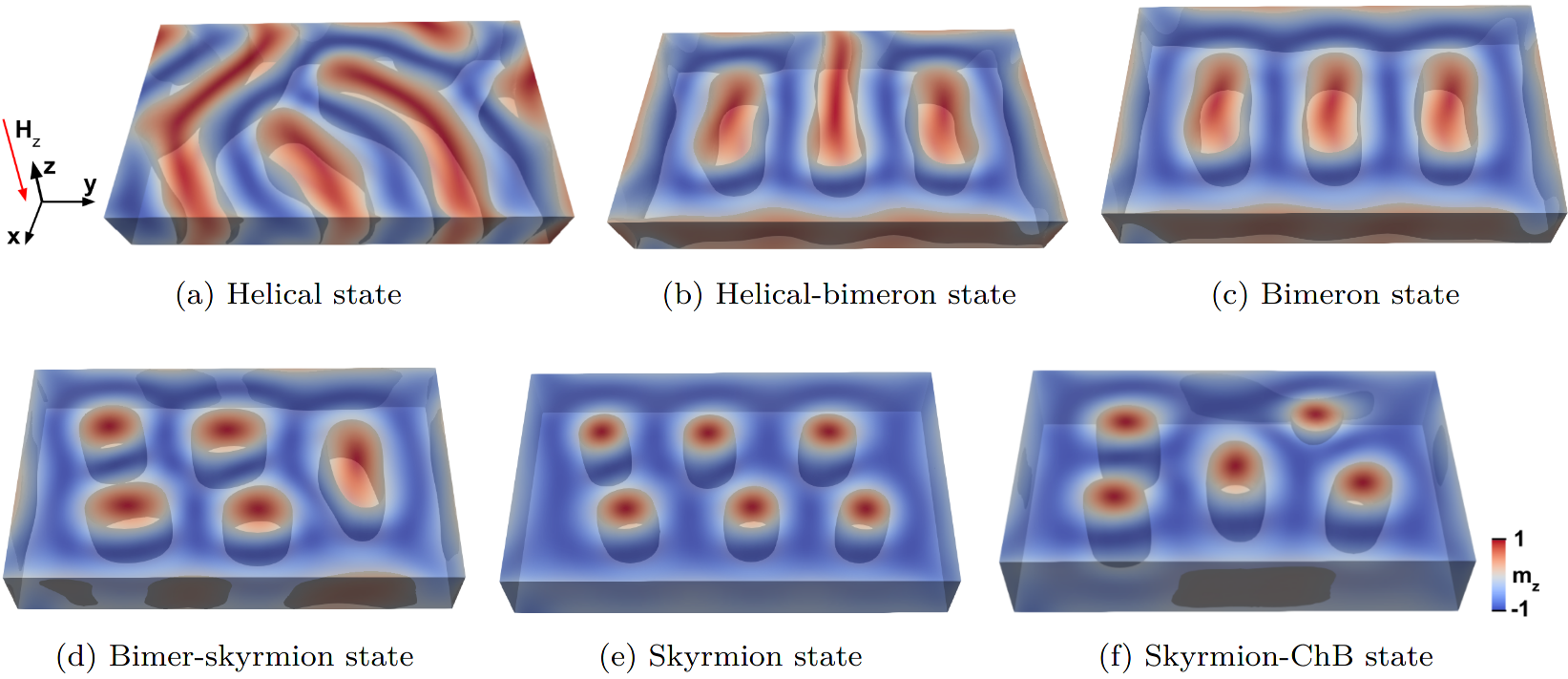}
\caption{\label{chiralStates}Non-trivial magnetization states forming in a rectangular FeGe platelet (\SI{310}{\nano\meter}$\times$\SI{180}{\nano\meter}) of varying thickness (between \SI{5}{\nano\meter} and \SI{60}{\nano\meter}) at different external field values. The color code describes the out-of-plane component $m_z$ of the normalized magnetization, and the isosurfaces indicate the regions where $m_z$ is equal to zero. Structures of this type appear at different film thicknesses as the external magnetic field is applied along the negative $z$ direction and is varied between \SI{0}{\milli\tesla} and \SI{900}{\milli\tesla}. The arrangement of these configurations in the image corresponds, roughly, to the order of preferential configurations appearing with increasing field strength.} 
\end{figure}

\unskip
We consider rectangular thin-film elements with a lateral size of $\SI{180}{\nano\meter}\times\SI{310}{\nano\meter}$ and thicknesses ranging between \SI{5}{\nano\meter} and \SI{75}{\nano\meter}, and simulate the magnetic structures forming in the presence of a perpendicular external magnetic field with a flux density varying between \SI{0}{\milli\tesla} and \SI{900}{\milli\tesla}.
The simulations yield a large variety of possible magnetization states in this thickness and field range, which can be classified into six non-trivial types, summarized in Fig.~\ref{chiralStates}. The three main types are the helical state shown in panel (a), which is characterized by the presence of regular spin spirals extending over large parts of the sample, the bimeron state (c), which can be interpreted either as particular a type of skyrmion structures that is stretched along one axis or, alternatively, as a helical state in which the extension of the helices is limited, and finally the skyrmion lattice state (e), which is characterized by a regular, hexagonal arrangement of skyrmions. 

In addition to these three fundamental states, there are also hybrid states in which two different types of structures coexists, such as the helical-bimeron state shown in Fig.~\ref{chiralStates}b) and the bimeron-skyrmion state shown in Fig.~\ref{chiralStates}d). These mixed states can be considered as transitional configurations between one fundamental state and another, which appear with changes in the external field value or in the film thickness. 
At elevated field values and at larger film thicknesses, a quasi-homogeneous state is formed (not shown), where the magnetization is largely aligned along the external field direction. At fields below saturation, chiral bobber (ChB) \cite{rybakov_new_2015} structures are also observed. These complex configurations of the magnetization can be considered as variants of skyrmions which do not traverse the entire thickness of the sample. Instead, they have a skyrmion-like structure only on one surface, which evolves into a quasi-saturated configuration on the opposite surface on a path along the film thickness. The apex of the ChB contains a Bloch point at which the magnetic structure changes in a discontinuous way. ChB structures have interesting micromagnetic properties and have recently been discussed as magnetic structures that could be attractive in the context of spintronic devices \cite{zheng}, but they are not of primary interest for our study. We display an example of a ChB structure only for completeness in the upper right of Fig.~\ref{chiralStates}f), where it coexists alongside four ordinary skyrmions. The magnetic structures were obtained by starting from a random initial configuration and by a subsequent energy minimization. For more details see section \ref{matmet}.

\subsection{Phase diagram of the magnetization states}
\begin{figure}[H]
\centering
\includegraphics[width=\textwidth]{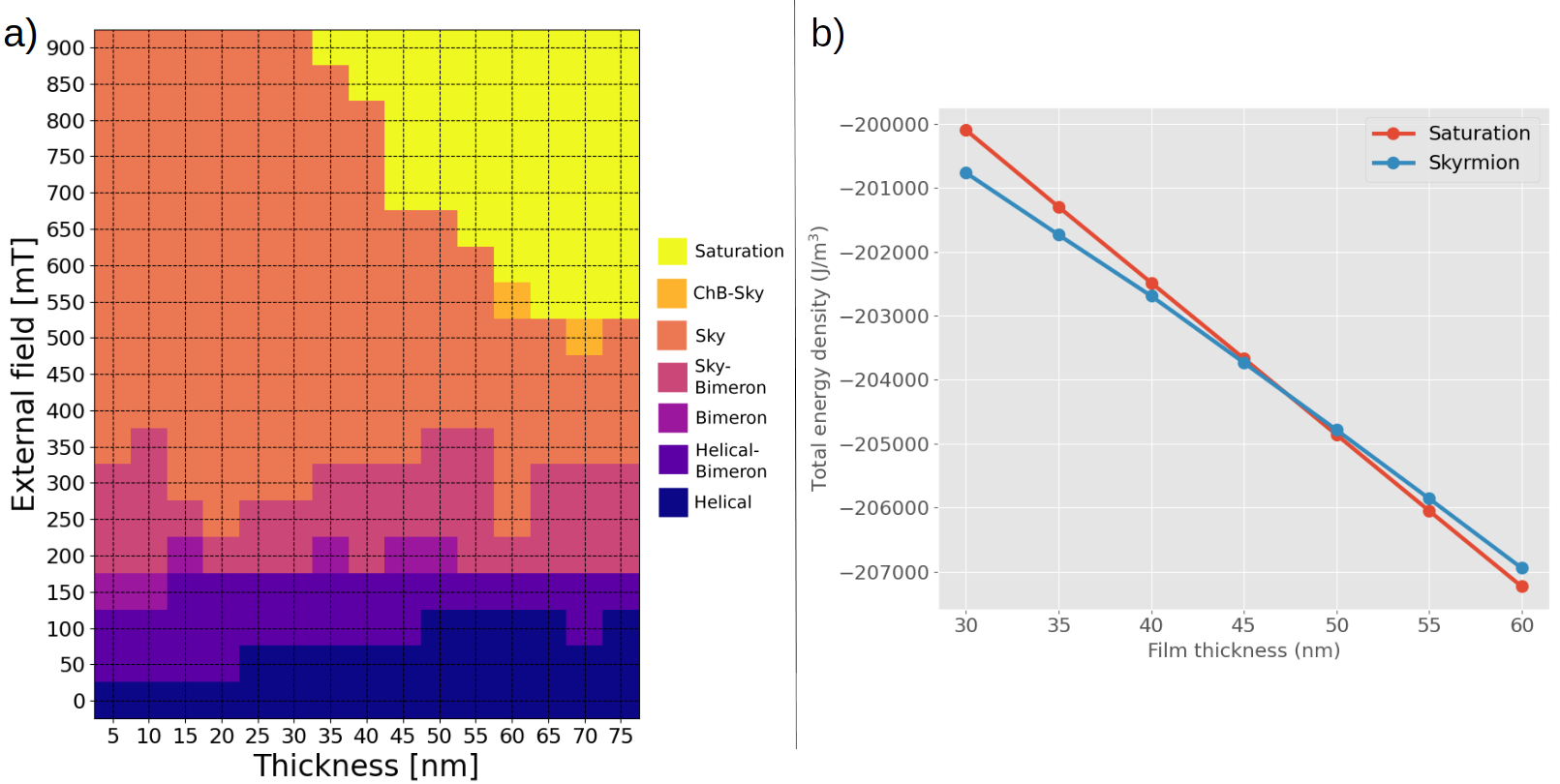}
\caption{\label{phaseDiag}a) Phase diagram displaying the lowest-energy magnetic configuration in the FeGe platelet as a function of the film thickness and the external field strength. At high fields and large film thickness, the sample is in a quasi-saturated state. By lowering the film thickness, the formation of skyrmion structures tends to become energetically favorable. b) Energy density of the skyrmions state (blue) and the quasi-saturated state (red) as a function of the film thickness at \SI{650}{\milli\tesla} field.} 
\end{figure}

The various magnetic states described in the previous section are possible equilibrium configurations of the magnetization forming in the FeGe platelets at different values of the external field and the film thickness. It is important to note that these magnetic structures are not uniquely determined by the film thickness and the field strength. Because of this, in order to avoid possible misunderstandings, we did not specify the  values of the thickness and the field strength at which the states shown in Fig.~\ref{chiralStates} occur. In fact, several metastable states that can be significantly different from each other are often possible under identical conditions, depending only on the magnetic history of the sample or, in a numerical experiment, on the initial conditions of the simulation. While it is generally not possible to identify a unique magnetization state that develops in the thin-film element, micromagnetic simulations can be used to determine the type of magnetic structure that has the lowest energy. The results of these calculations are summarized in the phase diagram shown in Fig.\ref{phaseDiag}a). 

Although the magnetic structure at a specific thickness and field value is generally not unique, the phase diagram helps identifying the most preferable structure as far as the total energy is concerned. While at lower field values (below about \SI{400}{\milli\tesla}) the phase diagram is rather complex, evidencing a multitude of possible magnetic structures showing neither any clearly dominating state nor a significant thickness dependence, the situation becomes simpler at larger field strengths (above about \SI{600}{\milli\tesla}). 
Two main states emerge in these ranges of larger field values: the skyrmion configuration and the quasi-saturated states. Moreover, these states are separated by a clearly defined boundary in the phase diagram, showing a distinct impact of the film thickness. Specifically, if a field of \SI{650}{\milli\tesla} is applied, the formation of skyrmion structures will be energetically favorable if the film thickness is below \SI{50}{\nano\meter}, while a quasi-saturated state will be the lowest-energy configuration at larger thickness values, as shown in Fig.~\ref{phaseDiag}b). 
This observation represents the fundamental of the concept of geometrically constrained skyrmions that we present in this study. The idea is the following. If the film thickness is {\em locally} modulated within a small dot-shaped region such that, at a given field, the skyrmion structure is favorable in that thinner part while in the rest of the sample the thickness is large enough to favor a quasi-homogeneous state, these thickness modulations can be designed to capture skyrmions. As we will show, this patterning makes it possible to generate pinning sites for skyrmions and, to some extent, to achieve a geometric control of the skyrmion position within the thin-film element.

\subsection{Geometrically constrained skyrmions}
\begin{figure}[H]
\centering
\includegraphics[width=\textwidth]{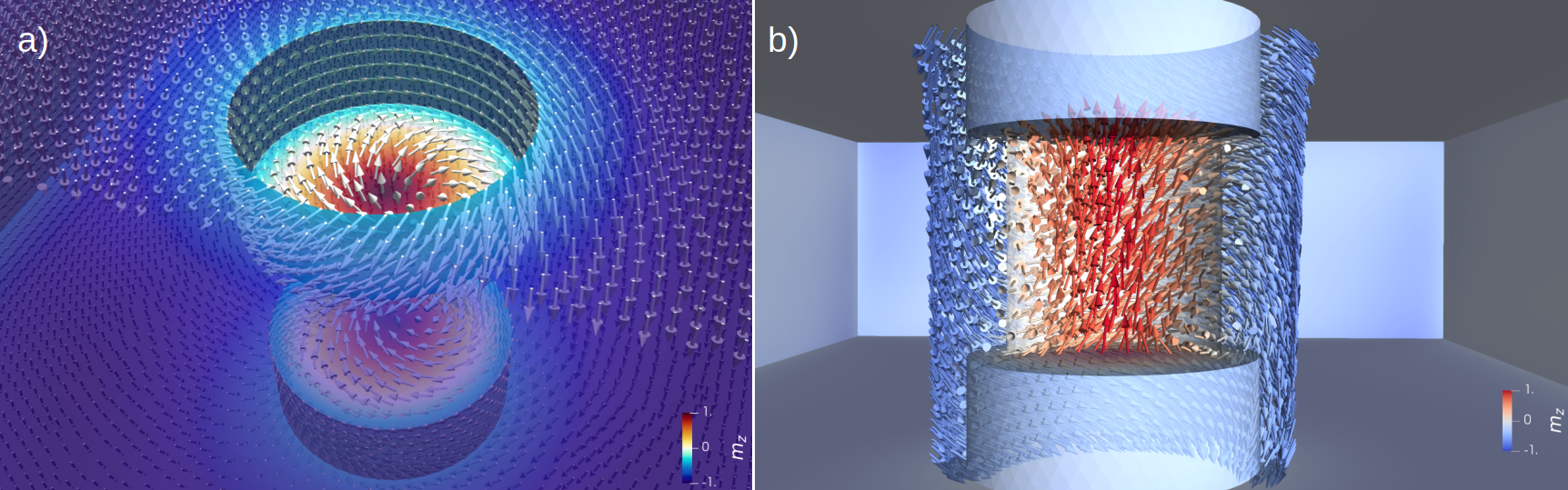}
\caption{\label{constrainedSky} a) A skyrmion is formed at the base of the cylindrical pocket. At the inner cylinder surface of the cavities, the magnetization circulates on closed loops, thereby facilitating the formation of the skyrmion in the center. The semitransparent representation of the surfaces shows the formation of the skyrmion in both pockets, on the top and the bottom surface. The magnetic structure is displayed by arrows on the sample surfaces. Some of the arrows have been remove in order to improve the visibility of the structure. b) View on the simulated skyrmion structure from inside the film. The skyrmion core connects the bases of the cylindrical pockets in the positive $z$ direction, while the surrounding volume is magnetized in the negative $z$ direction. The core of the skyrmion is delimited by a cylindrical isosurface $m_z=0$, shown here as a weak, transparent contrast in order to preserve the view on the central magnetic structure. Only a small subset of the computed arrows of the magnetization direction calculated within the volume is displayed.}
\end{figure}

We now consider magnetic structures forming in a FeGe platelet of \SI{60}{\nano\meter} thickness containing dot-like cylindrical cavities within which the thickness is locally reduced to \SI{30}{\nano\meter}. The phase diagram displayed in Fig.~\ref{phaseDiag} suggests that, at external field values of about \SI{650}{\milli\tesla}, the insertion of these cavities results in a geometry with specific regions favoring the stability of skyrmion structures in a thin-film element which, without such modulation, would tend to form a quasi-homogeneous magnetic configuration. This can lead to the formation, or the trapping, of skyrmions that are geometrically constrained to the regions in which the pockets have been introduced. Fig.~\ref{constrainedSky} shows such a geometrically constrained skyrmion in a \SI{60}{\nano\meter} thick platelet. The skyrmion remains confined to the small region in which the thickness is reduced by \SI{50}{\percent} through two cylindrical pockets with depth of \SI{15}{\nano\meter} and radius of $r=\SI{20}{\nano\meter}$, inserted symmetrically on both the top and the bottom surface of the film.

The geometrically constrained skyrmion, shown in Fig.~\ref{constrainedSky}, is stabilized by the geometry for two reasons. Firstly, as discussed before, in this field range the skyrmion state is generally favored because of the reduced film thickness. Secondly, the vortex-like magnetic configuration forming on the interior cylinder surfaces of the cavity helps pinning the position of the skyrmion to the center of the pocket. This cylindrical flux-closure structure thereby provides boundary conditions, albeit not in a mathematical sense, which constrain the skyrmion to this dot-like geometry. By forming such a cylindrical vortex structure, the magnetization finds a nearly optimal way to adapt to competing micromagnetic interactions. It thereby satisfies both the tendency of the DMI to introduce chiral, swirling patterns as well as the tendency imposed by the magnetostatic interaction to form flux-closure structures with the magnetization aligned along the surfaces. Without the geometric modification in the form of pockets on the surface, the magnetic structure would be in a quasi-homogeneous state. The simulations show that a symmetric insertion of these pockets on both the top and the bottom surfaces is necessary to obtain the desired stability and localization of skyrmions. If the thickness variation is introduced only on one of the surfaces, the pinning of skyrmions appears to be much less effective. 

If geometric modifications of the sample surface as described above can stabilize a skyrmion that would otherwise not form, the question arises whether this effect can be used to place skyrmions at specific positions where they might be generated or removed in a controlled way through external manipulation. This could be of interest, {\em e.g.}, for device concepts in which skyrmions are utilized as binary units of information, in a context similar to that of dot-patterned magnetic media for high-density data storage \cite{ross_patterned_2001}. In this case, the skyrmion pockets would take the role of the magnetic nanodots in bit-patterned media. While it is beyond the scope of this study to discuss the technical feasibility of such storage media or to explore the ability to write and delete individual skyrmion patterns into the pockets, we can show that, indeed, it is possible to stabilize skyrmions in various geometrically predefined locations that could be addressed individually.

\begin{figure}[H]
\centering
\includegraphics[width=\textwidth]{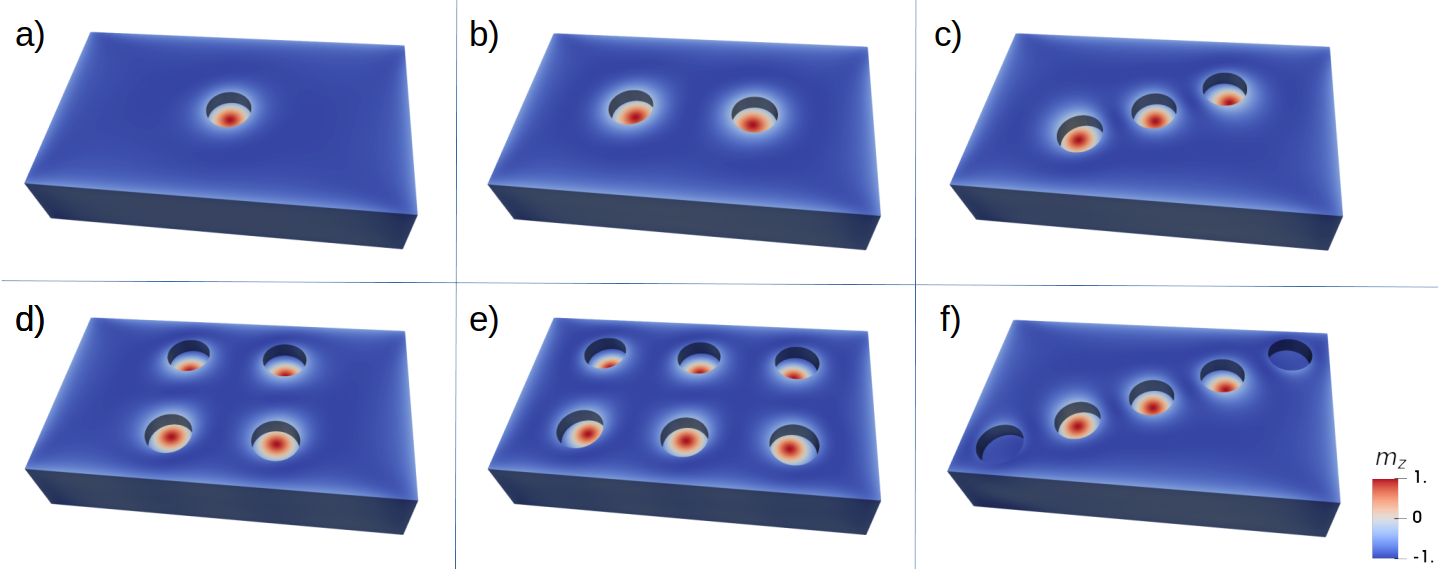}
\caption{\label{lego} Geometrically constrained skyrmions in FeGe platelets. By introducing circular pockets at specific positions, skyrmions can be artificially stabilized at positions that they would otherwise not attain. The geometric control, however, is not unlimited. Attempts to pack skyrmions too closely or to place them too close to the sample boundary can fail. This is shown in panel f), where skyrmions are stabilized only in the three central pockets, while the two outermost pockets remain empty.} 
\end{figure}

Fig.~\ref{lego})a-e) shows several examples of simulations in which the position of skyrmions in a thin-film element can be predetermined by introducing several pockets of the type discussed before. As shown in Fig.~\ref{lego}e), our simulations predict the possibility to stabilize six skyrmions at well-defined positions, placed on a regular grid, in our sub-micron FeGe platelet.
Although the results shown in Fig.~\ref{lego} may suggest a nearly optimal geometric control of the skyrmion positions, it is important to note that the pockets discussed here merely provide {\em preferential sites} for skyrmions. The latter may or may not form or remain pinned at those sites. In particular, it is not sufficient to thin-out a part of the sample in a sample to ensure the appearance of geometrically constrained skyrmions. The purpose of such pockets could rather be to capture existing skyrmions and to fix their positions at well-defined positions, similar to the domain-wall pinning role that is played by notches in conventional racetrack-memory devices \cite{hayashi_dependence_2006,parkin_magnetic_2008}. It should also be noted that the geometric trapping of skyrmions with such pockets does not always work, in particular when the pockets are too closely packed. As a rule of thumb, the material must observe a characteristic minimal distance between the skyrmions that is given by the material-dependent long-range helical period $l_D$, which in the case of FeGe is about \SI{70}{\nano\meter} (see section \ref{matmet}). We also found that skyrmions cannot be stabilized at positions too close to the lateral sample boundaries due to repulsion \cite{brearton2020magnetic}. An example of such a failed attempt is shown in Fig.~\ref{lego}f). 
In spite of these limitations, the ability to geometrically constrain skyrmions provides an attractive way to obtain control over the skyrmion position in thin-film elements, which could have important technological implications.

\section{Discussion}
By means of micromagnetic finite-element simulations we have presented a possibility to control the position of magnetic skyrmions at predefined positions within a thin-film element by introducing cylindrical nano-pockets graved into the surface. Our concept to geometrically constrain skyrmions {\em via} such dot-like thickness variations is in many ways analogous to the idea of geometrically constrained domain walls \cite{bruno_geometrically_1999} in cylindrical nanowires, or to studies in which indentations have been introduced in rectangular strips in order to capture head-to-head domain walls in race-track type memory devices \cite{bogart_effect_2008}. In those cases, too, the desired effect of the geometric constraint is to define preferential sites for specific micromagnetic structures, such that the magnetic structures constrained at those artificial pinning sites require a certain activation energy in order to detach from them. The pockets described in this work could effectively play this role in the case of skyrmions driven along magnetic strips by means of spin-polarized electrical currents. Such a geometric control of their position would allow shifting skyrmions between well-defined points on the track. Moreover, as mentioned before, the trapping of skyrmions at dot-like sites could also serve as a principle for skyrmion-based data storage devices, without necessarily involving any displacement or depinning processes. If skyrmions can be selectively generated and dissolved at such preferential sites, {\em e.g.}, by means of the field of a magnetized nano-tip or through a localized spin-polarized current traversing the film thickness, the geometrically constrained skyrmions could represent units of information that could be written and erased. Perhaps such skyrmionic dot material could even be stacked in three dimensions for ultra-high density storage purposes. To address such possibilities, future research directions could explore ways to reversibly insert skyrmions in these geometrically defined regions. Another potentially interesting use of our concept concerns magnonic applications \cite{garst2017collective}. Since skyrmions can act as point-like scattering centers for spin waves, the ability to arrange them at specific sites as described in this study could open up new perspectives, as this could result in a new type of magnonic metamaterials in the form of artificial magnon Bragg lattices consisting of skyrmions arranged on regular lattice sites. Such artificial structures could be tailored to yield specific scattering and interference properties for spin waves that could not be obtained otherwise.

\section{Materials and Methods\label{matmet}}
The material modelled in this study is FeGe. Due to its well-known helimagnetic properties, this B20-type non-centrosymmetric material serves as a prototype for materials hosting chiral magnetic structures that develop due to the ``bulk'' DMI effect, as opposed to certain systems of ultrathin magnetic films and substrates that can generate an ``interfacial'' DMI \cite{finocchio2016magnetic}. The micromagnetic parameters of FeGe are \cite{beg_ground_2015} $A=\SI{8.78e-12}{\joule\per\meter}$, $M_{\text s}=\SI{384}{\kilo\ampere\per\meter}$, and $D=\SI{1.58e-3}{\joule\per\meter\squared}$, where $A$ is the ferromagentic exchange constant, $M_{\text s}$ the saturation magnetization and $D$ the DMI constant. We neglect any magnetocrystalline anisotropy of the material, setting the uniaxial anisotropy to zero, $K_{\text u} = \SI{0}{\joule\per\meter\cubed}$. A characteristic length scale of this material is the long-range helical period $l_d=4\pi A/\left|D\right|\simeq\SI{70}{\nano\meter}$. This length scale describes the typical period length of magnetic spirals forming as a result of the competing interactions of the ferromagnetic exchange on one hand and the DMI on the other.

With these parameters, the total energy $E_\text{tot}$ of the system is given by the sum of the Zeeman term, the ferromagnetic exchange, the DMI interaction and the magnetostatic energy:

\begin{equation}
E_\text{tot} = \int\limits_{V} \left( \mu_0\bm{H}_\text{ext}\cdot\bm{M} + A\cdot\sum\limits_{i=x,y,z}\left(
\bm{\nabla}\bm{m}_i\right)^2 + D\bm{m}\cdot\left(\bm{\nabla}\times\bm{m}\right) - \frac{\mu_0}{2}\bm{M}\cdot\bm{\nabla}u \right)
\,\text{d}V
\end{equation}

Here $V$ is the sample volume, $\bm{H}_{\text ext}$ is the externally applied magnetic field, $\mu_0=4\pi\times10^{-7}\si{\volt\second\per\ampere\per\meter}$ is the vacuum permeability, $\bm{m}=\bm{M}/M_\text{s}$ is the reduced (normalized) magnetization, and $u$ is the magnetostatic scalar potential.
The  magnetostatic (demagnetizing) field $\bm{H}_\text{d} = -\bm{\nabla}u$ is the gradient field of the magnetostatic potential. We calculate the magnetostatic potential $u$, which accounts for the long-range dipolar interaction, by using the hybrid finite-element method / boundary element method (FEM/BEM) introduced by Fredkin and Koehler \cite{fredkin1990hybrid,koehler1992finite}. The dense matrix occurring in the  boundary integral part of this formalism is represented using ${\cal H}^2$ type hierarchical matrices \cite{hertel_large-scale_2019}. This data-sparse representation effectively overcomes size limitations arising from the boundary element method, as it yields a {\em linear} scaling of the computational resources required for the calculation of the magnetostatic term, which would otherwise grow {\em quadratically} with the number of degrees of freedom on the surface.

For each energy term, an effective field $\bm{H}_\text{eff}$ is defined as the variational derivative of the corresponding partial energy $E$, 
\begin{equation}
\bm{H}_\text{eff}(\bm{r},t)=-\frac{\delta E\left[\bm{M}\left(\bm{r},t\right)\right]}{\mu_0\delta\bm{M}}
\end{equation}

Specifically, the effective field of the ferromagnetic exchange is 
\begin{equation}
\mu_0\bm{H}^{(\text{xc})}_\text{eff}(\bm{r},t) = -2A\Delta\bm{m}
\end{equation}
and the effective field of the DMI is 
\begin{equation}
\mu_0\bm{H}^{(\text{DMI})}_\text{eff}(\bm{r},t) = -2D\left(\nabla\times\bm{m}\right)
\end{equation}

Together with the magnetostatic field and the external (Zeeman) field, these effective fields enter the Landau-Lifshitz-Gilbert (LLG) equation \cite{gilbert2004phenomenological}, which describes the evolution of the magnetization field $\bm{M}(\bm{r},t)$ in time,
\begin{equation}
\label{llg}\frac{\text{d}\bm{M}}{\text{d}t} = -\gamma\left(\bm{M}\times\bm{H}_\text{eff}\right)+\frac{\alpha}{M_s}\left(
\bm{M}\times\frac{\text{d}\bm{M}}{\text{d}t}\right)
\end{equation}
where $\gamma$ is the gyromagnetic ratio and $\alpha$ is a phenomenological, dimensionless damping constant. We use the LLG equation to calculate equilibrium structures $\bm{M}(\bm{r})$ of the magnetization, by integrating in time until convergence is reached. In the numerical simulations, convergence is achieved when either the total energy ceases to change over a long period, or when the torque (magnitude of the right hand side of the LLG equation) drops below a user-defined threshold. 

The geometry of the samples is designed with FreeCAD \cite{freecad} and the discretization into linear tetrahedral elements is performed with Netgen \cite{schoberl1997netgen}. The visalization of the FEM data was done with ParaView \cite{paraview}. The cell size does not exceed \SI{2.5}{\nano\meter}, which is well below the exchange length $l_\text{ex}=\sqrt{2A/\mu_0M_s^2}\simeq\SI{9.7}{\nano\meter}$, in order to avoid discretization errors. The finite element meshes in this study contain typically about one million of finite elements. The micromagnetic simulations are done with our proprietary GPU-accelerated finite-element software \cite{hertel_large-scale_2019}. 

The discretized representation of the vector field of the magnetization is given by a value of $M_i$ defined at each node (vertex) $i$ of the finite-element mesh. The magnetostatic field as well as the effective fields of the ferromagnetic exchange and the DMI are calculated within in each tetrahedral element. The element-based data of these fields is then mapped onto the nodes of the mesh, in order to calculate the effective field acting on the magnetization and thus to calculate the evolution of the magnetization in time at each node according to the LLG equation. More details on the calculation of the converged magnetization states are given in appendix \ref{emin}.

\vspace{6pt}

\authorcontributions{Conceptualization, S.A.P; methodology, R.H. and S.A.P.; software, R.H.; writing--original draft preparation, R.H. and S.A.P.; writing--review and editing, R.H. and S.A.P.; visualization, R.H. and S.A.P.; supervision, R.H.; project administration, R.H.; funding acquisition, R.H. All authors have read and agreed to the published version of the manuscript.}

\funding{
This work has benefited from support by the initiative of excellence IDEX-Unistra (ANR-10-IDEX-0002-02) through the French National Research Agency (ANR) as part of the ``Investment for the Future'' program.
}

\acknowledgments{
The authors acknowledge the High Performance Computing center of the University of Strasbourg for supporting this work by providing access to computing resources. Part of the computing resources were funded by the Equipex Equip@Meso project (Programme Investissements d'Avenir) and the CPER Alsacalcul/Big Data.

}

\conflictsofinterest{The authors declare no conflict of interest. The funders had no role in the design of the study; in the collection, analyses, or interpretation of data; in the writing of the manuscript, or in the decision to publish the results.}

\abbreviations{The following abbreviations are used in this manuscript:\\

\noindent 
\begin{tabular}{@{}ll}
DMI & Dzyaloshinksii-Moryia interaction\\
LLG & Landau-Lifshitz-Gilbert equation\\
FEM & Finite Element Method \\
BEM & Boundary Element Method \\
GPU & Graphical Processing Unit\\
3D & three-dimensional\\
ChB & chiral bobber 
\end{tabular}}

\appendixtitles{no} %
\appendix
\section{Energy minimization\label{emin}}
\unskip
Because in this particular study we are not interested in the dynamic evolution of the magnetization but only in static, converged magnetic structures, the integration of the LLG equation in the code fulfils the practical role of guiding the system  along a path of energy-minimization in an iterative way. Since the dynamics of the magnetization during the transition from the initial to the converged state is irrelevant for this work, we are free to choose a conveniently large damping parameter $\alpha = 0.5$ in order to accelerate the energy minimization. Furthermore, we remove the precession term by setting $\gamma=0$ in eq.~(\ref{llg}), thereby effectively using a damped LaBonte-type energy minimization scheme instead of following the path of the magnetization dynamics described by the LLG equation. The numerical integration is done with a Dormand-Prince algorithm \cite{ahnert_odeint_2011}, and the effective field values are refreshed several times during each time step. The choice of a large value of the damping allows us to use time steps of up to \SI{1}{\pico\second}, which is about ten times larger than the step size that we would usually employ in dynamic simulations with low damping. With these parameters, and owing to the GPU acceleration of our code, it takes only a short time (between several minutes and a few hours) to simulate the magnetic structures discussed in this work. To calculate the skyrmion states, we saturate the magnetization along the positive $z$ direction and subsequently let the system relax in the presence of an external magnetic field aligned along the negative $z$ direction. The $z$ axis is oriented parallel to the surface normal, as shown in Fig.~\ref{chiralStates}.

\reftitle{References}

\bibliography{constrainedSky_arxiv.bib}

\end{document}